# JOSEPHSON EFFECT IN POINT CONTACTS BETWEEN TWO-BAND SUPERCONDUCTORS

ALEXANDER OMELYANCHOUK, YURIY YERIN*

*B.Verkin Institute for Low Temperature Physics and Engineering of the National Academy of Sciences of Ukraine, 47 Lenin Ave., Kharkov 61103, Ukraine*

**Abstract.** The microscopic theory of Josephson effect in point contacts between two-band superconductors is developed. The general expression for the Josephson current, which is valid for arbitrary temperatures, is obtained. We considered the dirty superconductors with interband scattering, which produces the coupling of the Josephson currents between different bands. The influence of phase shifts and interband scattering rates in the banks is analyzed near critical temperature $T_c$. It is shown that for some values of parameters the critical current can be negative, which means the $\pi$-junction behavior.

## 1. Introduction

Discovery of high-temperature superconductivity in iron-based compounds[1] have expanded a range of multiband superconductors besides well-known magnesium diboride $MgB_2$ with $T_c$=39 K [2]. Two-band superconductivity proposes new interesting physics. In a large number researches the specific effects in temperature behavior of the first and upper critical fields[3-5] and London penetration depth[3,6,7] were demonstrated. Mixed state and peculiar vortex core structure were studied in[8]. The coexistence of two distinctive order parameters $|\Psi_1| = \Delta_1 \exp(i\phi_1)$ and $\Psi_2 = |\Delta_2|\exp(i\phi_2)$ renewed interest in phase coherent effects in superconductors. In the case of two order parameters the question arises, what is the phase shift $\phi_1 - \phi_2$ between $\Psi_1$ and $\Psi_2$? From the minimization of free energy it follows that in homogeneous equilibrium state this phase shift is fixed to 0 or π, depending on the sign of interband coupling. The phases $\phi_1$ and $\phi_2$ can be decoupled as small plasmon oscillations (Leggett mode[9]) or due to formation of phase slips textures in strong electric field[10]. The coherent current states and depairing curves have been calculated in[11], where it was shown the possibility of phase shift switching in homogeneous current state with increasing of the superfluid velocity $v_s$. Such switching manifests itself in the dependence $j(v_s)$ and also in Little-Parks effect[12].

---

*\* To whom correspondence should be addressed. Yuriy Yerin, B.Verkin Institute for Low Temperature Physics and Engineering of the National Academy of Sciences of Ukraine, 47 Lenin Ave., Kharkov 61103, Ukraine; e-mail: yerin@ilt.kharkov.ua*

The Josephson effect in superconducting junctions is the probe for research of phase coherent effects. The stationary Josephson effect in tunnel $S_1$-I-$S_2$ junctions (I - dielectric) between two- and one- band superconductors have been studied recently in a number of articles[13-15]. Another basic type of Josephson junctions are the junctions with direct conductivity, S-C-S contacts (C – constriction). As was shown in[16,17,18] the Josephson behavior of S-C-S structures qualitatively differ from properties of tunnel junctions. In this paper we generalize KO theory[16,17] of stationary Josephson effect in S-C-S point contacts for the case of two-band superconductors. Within the microscopic Usadel equations we calculate the Josephson current and study its dependence on the mixing of order parameters due to interband scattering and phase shifts in the contacting two-band superconductors.

## 2. Model and basic equations

Consider the weak superconducting link as a thing filament of length $L$ and diameter $d$, connecting two superconducting banks (Fig.1). Such model describes the S-C-S (Superconductor-Constriction-Superconductor) contacts with direct conductivity (point contacts, microbridges), which qualitatively differ from the tunnel S-I-S junctions. On condition that $d \ll L$ and $d \ll \min[\xi_1(0), \xi_2(0)]$ ($\xi_i(T)$ - coherence lengths) we can solve inside the filament ($0 \leq x \leq L$) a one-dimensional problem with "rigid" boundary conditions. At $x = 0, L$ all functions are assumed equal to the values in homogeneous no-current state of corresponding bank.

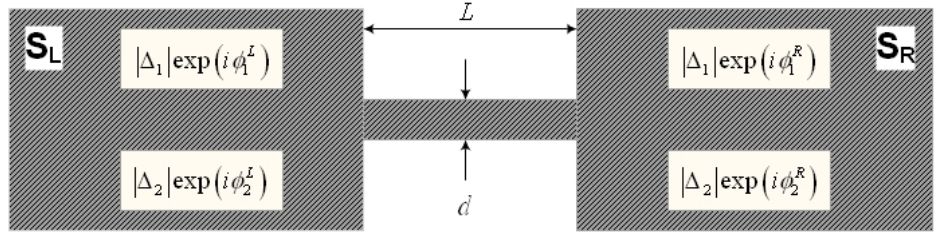

*Figure 1*. A model of Superconductor-Constriction-Superconductor contact (S-C-S contact). The right and left banks are massive two-band superconductors connected by the thing filament of length *L* and diameter *d*.

We investigate a case of two-band superconductor with strong impurity intraband scattering rates (dirty limit) and weak interband scattering. In the dirty limit superconductor is described by the Usadel equations[19] for normal and anomalous Green's functions $g$ and $f$, which for two-band superconductor take the form[3]:

$$\omega f_1 - D_1 \left( g_1 \nabla^2 f_1 - f_1 \nabla^2 g_1 \right) = \Delta_1 g_1 + \gamma_{12} \left( g_1 f_2 - g_2 f_1 \right), \qquad (1)$$

$$\omega f_2 - D_2 \left( g_2 \nabla^2 f_2 - f_2 \nabla^2 g_2 \right) = \Delta_2 g_2 + \gamma_{21} \left( g_2 f_1 - g_1 f_2 \right). \qquad (2)$$

Usadel equations are supplemented with self-consistency equations for order parameters $\Delta_i$:

$$\Delta_i = 2\pi T \sum_j \sum_{\omega>0}^{\omega_D} \lambda_{ij} f_j, \qquad (3)$$

and with expression for the current density

$$j = -ie\pi T \sum_i \sum_{\omega}^{\omega_D} N_i D_i \left( f_i^* \nabla f_i - f_i \nabla f_i^* \right). \qquad (4)$$

Index $i = 1,2$ numerates the first and second bands. Green's functions $g_i$ and $f_i$ are connected by normalization condition $g_i^2 + |f_i|^2 = 1$ and depend on $x$ and the Matsubara frequency $\omega = (2n+1)\pi T$. $D_i$ are the intraband diffusivities due to nonmagnetic impurity scattering, $N_i$ are the density of states on the Fermi surface for the electrons of the i-th band, electron-phonon constants $\lambda_{ij}$ take into account Coulomb pseudopotentials and $\gamma_{ij}$ are the interband scattering rates. There are the symmetry relations $\lambda_{12}N_1 = \lambda_{21}N_2$ and $\gamma_{12}N_1 = \gamma_{21}N_2$.

In considered case of short weak link $L \ll \min[\xi_1(0), \xi_2(0)]$ we can neglect all terms in the Eqs. (1), (2) except the gradient one. And using the normalization condition we have equations for $f_{1,2}$

$$\sqrt{1-|f_1|^2}\frac{d^2}{dx^2}f_1 - f_1\frac{d^2}{dx^2}\sqrt{1-|f_1|^2} = 0, \tag{5}$$

$$\sqrt{1-|f_2|^2}\frac{d^2}{dx^2}f_2 - f_2\frac{d^2}{dx^2}\sqrt{1-|f_2|^2} = 0. \tag{6}$$

The boundary conditions for Eqs. (5) and (6) are determined by solutions of equations for Green's functions in the banks:

$$\begin{cases} \omega f_1^{L(R)} = \Delta_1^{L(R)}\sqrt{1-\left|f_1^{L(R)}\right|^2} + \gamma_{12}\left(\sqrt{1-\left|f_1^{L(R)}\right|^2}f_2^{L(R)} - \sqrt{1-\left|f_2^{L(R)}\right|^2}f_1^{L(R)}\right), \\ \omega f_2^{L(R)} = \Delta_2^{L(R)}\sqrt{1-\left|f_2^{L(R)}\right|^2} + \gamma_{21}\left(\sqrt{1-\left|f_2^{L(R)}\right|^2}f_1^{L(R)} - \sqrt{1-\left|f_1^{L(R)}\right|^2}f_2^{L(R)}\right). \end{cases} \tag{7}$$

Introducing the phases of order parameters in banks

$$\Delta_1^{L(R)} = |\Delta_1|\exp(i\phi_1^{L(R)}), \Delta_2^{L(R)} = |\Delta_2|\exp(i\phi_2^{L(R)}), \tag{8}$$

and writing $f_i(x)$ in Eqs. (5) and (6) as $f_i(x) = |f_i(x)|\exp(i\chi_i(x))$ we have

$$|f_i(0)| = |f_i|, |f_i(L)| = |f_i|, \chi_i(0) = \chi_i^L, \chi_i(L) = \chi_i^R, \tag{9}$$

where $|f_i|$ and $\chi_i^{L(R)}$ are connected with $|\Delta_i|$ and $\phi_i^{L(R)}$ through the Eq. (7).

The solution of Eqs. (5)-(9) determines the Josephson current in the system. It depends on the phase difference on the contact $\phi \equiv \phi_1^R - \phi_1^L = \phi_2^R - \phi_2^L$ and from possible phase shifts in each banks $\delta\phi^L = \phi_1^L - \phi_2^L$ and $\delta\phi^R = \phi_1^R - \phi_2^R$. The phase shift $\delta\phi$ between the phases of the two order parameters in two-band superconductor can be 0 or $\pi$, depending on the sign on the interband coupling constants[20] and the values of the interband scattering rates.

Eqs. (5), (6) with boundary conditions (Eq. 9) admit simple analytical solution, and for the current (Eq. 4) we obtain

$$j = \frac{4e\pi T}{L} \sum_{\omega}^{\omega_D} N_1 D_1 \frac{|f_1| \cos \frac{\chi_1^L - \chi_1^R}{2}}{\sqrt{\left(1-|f_1|^2\right)\sin^2 \frac{\chi_1^L - \chi_1^R}{2} + \cos^2 \frac{\chi_1^L - \chi_1^R}{2}}}$$

$$\times \arctan \frac{|f_1| \sin \frac{\chi_1^L - \chi_1^R}{2}}{\sqrt{\left(1-|f_1|^2\right)\sin^2 \frac{\chi_1^L - \chi_1^R}{2} + \cos^2 \frac{\chi_1^L - \chi_1^R}{2}}}$$

$$+ \frac{4e\pi T}{L} \sum_{\omega}^{\omega_D} N_2 D_2 \frac{|f_2| \cos \frac{\chi_2^L - \chi_2^R}{2}}{\sqrt{\left(1-|f_2|^2\right)\sin^2 \frac{\chi_2^L - \chi_2^R}{2} + \cos^2 \frac{\chi_2^L - \chi_2^R}{2}}}$$

$$\times \arctan \frac{|f_2| \sin \frac{\chi_2^L - \chi_2^R}{2}}{\sqrt{\left(1-|f_2|^2\right)\sin^2 \frac{\chi_2^L - \chi_2^R}{2} + \cos^2 \frac{\chi_2^L - \chi_2^R}{2}}}. \quad (10)$$

This general expression together with Eqs. (7) describes the Josephson current as function of gaps in the banks $|\Delta_i|$ and phase difference on the contact $\phi \equiv \phi_1^R - \phi_1^L = \phi_2^R - \phi_2^L$. If we neglect the interband scattering $\gamma_{ik}$, the equations (7) for $f_{1,2}$ become decoupled and the current (Eq. 10) consists of two independent inputs from transitions $1 \rightarrow 1$ and $2 \rightarrow 2$. Thus in this case we have for each components the Josephson currents in KO theory with corresponding values of $\Delta_{1,2}$. The interesting case is the mixing of different contributions due to the interband scattering. For arbitrary values of $\gamma_{ik}$ and arbitrary temperature $T$ Eqs. (7) can be solved numerically. To study the effects of interband scattering we consider the Josephson current near critical temperature $T_c$.

### 3. Josephson current near critical temperature $T_c$

Near $T_c$ Eqs. (5)-(7) are considerably simplified and we have equations for $f_{1,2}(x)$

$$\frac{d^2}{dx^2} f_1 = 0, \quad \frac{d^2}{dx^2} f_2 = 0, \quad (11)$$

with boundary conditions

$$f_1^{L(R)} = \frac{(\omega + \gamma_{21})|\Delta_1|\exp(i\phi_1^{L(R)}) + \gamma_{12}|\Delta_2|\exp(i\phi_2^{L(R)})}{\omega^2 + (\gamma_{12} + \gamma_{21})\omega},$$
$$f_2^{L(R)} = \frac{(\omega + \gamma_{12})|\Delta_2|\exp(i\phi_2^{L(R)}) + \gamma_{21}|\Delta_1|\exp(i\phi_1^{L(R)})}{\omega^2 + (\gamma_{12} + \gamma_{21})\omega}.$$

(12)

The current density (Eq. 4) with solutions of Eqs. (11) takes the form:

$$j = -\frac{ie\pi T}{L}\sum_i \sum_\omega^{\omega_D} N_i D_i \left( \left(f_i^L\right)^* f_i^R - f_i^L \left(f_i^R\right)^* \right),$$

(13)

where functions $f_i^{L(R)}$ are related to the order parameters in the banks by the expressions (Eqs. 12).

The current density $j$ (Eq. 13) consists of four partials inputs produced by transitions from left bank to right bank between different bands

$$\begin{aligned} j_{11} &\sim |\Delta_1|^2 \sin(\phi_1^R - \phi_1^L), \\ j_{22} &\sim |\Delta_2|^2 \sin(\phi_2^R - \phi_2^L), \\ j_{12} &\sim |\Delta_1||\Delta_2| \sin(\phi_2^R - \phi_1^L), \\ j_{21} &\sim |\Delta_1||\Delta_2| \sin(\phi_1^R - \phi_2^L). \end{aligned}$$

(14)

The relative directions of components $j_{ik}$ depend on the intrinsic phase shifts in the banks $\delta\phi^{L,R}$ (Fig.2).

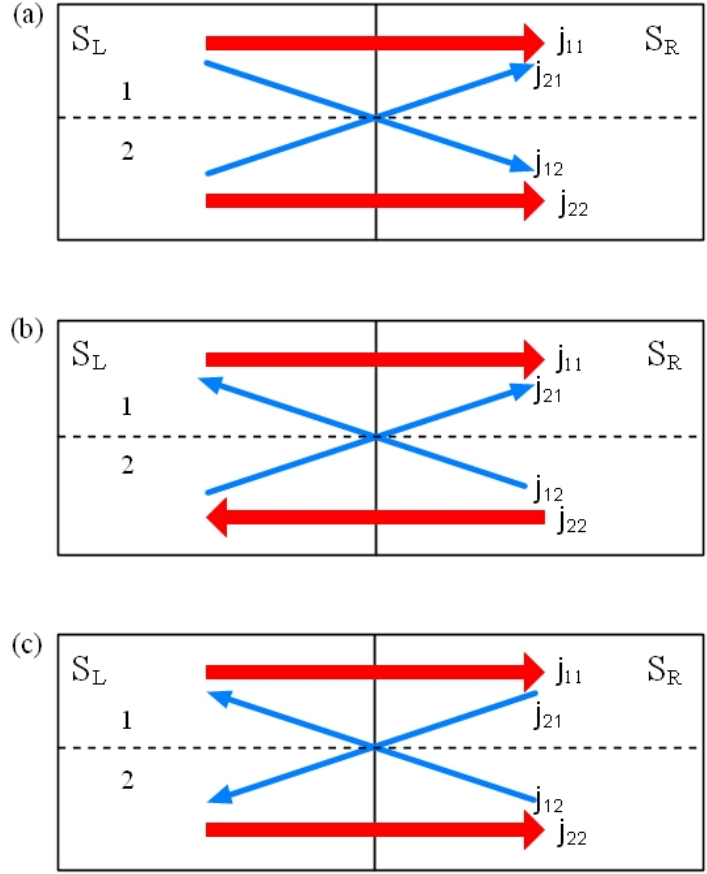

*Figure 2*. Current directions in S-C-S contact between two-band superconductors. (a) – there is no shift between phases of order parameters in the left and right superconductors; (b) – π-shift is present in the right superconductor and is absent in the left superconductor; (c) - there is the π-shift of order parameters phases at the both banks.

Introducing resistance of contact in a normal state:

$$\frac{1}{R_N} = \frac{2Se^2}{L}(N_1 D_1 + N_2 D_2), \qquad (15)$$

where $S$ is the contact cross-section area, for the current components we have
1. For $\delta\phi^L = 0$ and $\delta\phi^R = 0$:

$$I = \frac{\pi T}{eR_N (N_1 D_1 + N_2 D_2)} \left( N_1 D_1 \sum_{\omega}^{\omega_D} \frac{(|\Delta_1|(\omega + \gamma_{21}) + |\Delta_2|\gamma_{12})^2}{(\omega^2 + (\gamma_{12} + \gamma_{21})\omega)^2} + N_2 D_2 \sum_{\omega}^{\omega_D} \frac{(|\Delta_2|(\omega + \gamma_{12}) + |\Delta_1|\gamma_{21})^2}{(\omega^2 + (\gamma_{12} + \gamma_{21})\omega)^2} \right) \sin\phi = I_c \sin\phi,$$

(16)

2. For $\delta\phi^L = \pi$ and $\delta\phi^R = \pi$:

$$I = \frac{\pi T}{eR_N (N_1 D_1 + N_2 D_2)} \left( N_1 D_1 \sum_{\omega}^{\omega_D} \frac{(|\Delta_1|(\omega + \gamma_{21}) - |\Delta_2|\gamma_{12})^2}{(\omega^2 + (\gamma_{12} + \gamma_{21})\omega)^2} + N_2 D_2 \sum_{\omega}^{\omega_D} \frac{(|\Delta_2|(\omega + \gamma_{12}) - |\Delta_1|\gamma_{21})^2}{(\omega^2 + (\gamma_{12} + \gamma_{21})\omega)^2} \right) \sin\phi = I_c \sin\phi,$$

(17)

3. For $\delta\phi^L = 0$ and $\delta\phi^R = \pi$:

$$I = \frac{\pi T}{eR_N (N_1 D_1 + N_2 D_2)} \left( N_1 D_1 \sum_{\omega}^{\omega_D} \frac{\Delta_1^2(\omega + \gamma_{21})^2 - \Delta_2^2 \gamma_{12}^2}{(\omega^2 + (\gamma_{12} + \gamma_{21})\omega)^2} + N_2 D_2 \sum_{\omega}^{\omega_D} \frac{\Delta_1^2 \gamma_{21}^2 - \Delta_2^2(\omega + \gamma_{12})^2}{(\omega^2 + (\gamma_{12} + \gamma_{21})\omega)^2} \right) \sin\phi = I_c \sin\phi,$$ (18)

From expressions (Eq. 18) it follows that at certain values of constants $\gamma_{12,21}$, ratio $N_2 D_2 / N_1 D_1$ and of the values of gaps $|\Delta_1|$ and $|\Delta_2|$ the critical current $I_c$ can be negative. It means that in this case we have the so-called π-junction[21,22] (see illustration in Fig.3).

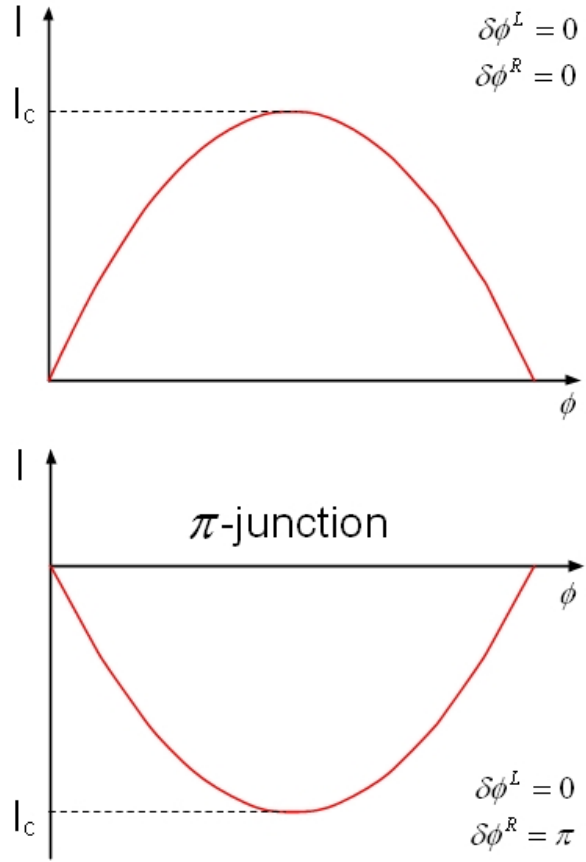

*Figure 3*. Current-phase relations for different phase shifts in the banks.

## 4. Conclusion

The microscopic theory of Josephson effect in point contacts between two-band superconductors is developed. The general expression for the Josephson current, which is valid for arbitrary temperatures, is obtained. We considered the dirty superconductors with interband scattering. Interband scattering in the contacting superconductors produces the coupling of the currents between different bands. The influence of phase shifts and interband scattering rates in the banks is analyzed near critical temperature $T_c$. It is shown that for some values of parameters the critical current can be negative, which means the $\pi$-junction behavior.

We acknowledge partial support from the FRSF (grant F28.21019) and NASU program "nanostructures, nanomaterials and nanotechnologies".